\documentclass[3p,11pt,a4paper,sort&compress]{elsarticle}
\usepackage{amsmath,amssymb,ascmac,comment,color}
\usepackage[utf8]{inputenc}
\usepackage[colorlinks=true,urlcolor=blue,anchorcolor=blue,citecolor=blue,linkcolor=blue,filecolor=blue,menucolor=blue,pagecolor=blue,linktocpage=true,pdfproducer=medialab,pdfa=true]{hyperref}

\begin{document}

\title{Note on Covariant St\"{u}ckelberg Formalism and Absence of
Boulware-Deser Ghost in Bi-gravity}

\author{Toshifumi Noumi}
\ead{iasnoumi@ust.hk}

\address{Jockey Club Institute for Advanced
Study, Hong Kong University of Science and Technology, Hong Kong}

\author{Masahide Yamaguchi}
\ead{gucci@phys.titech.ac.jp}

\author{Daisuke Yoshida}
\ead{yoshida@th.phys.titech.ac.jp}

\address{Department of Physics, Tokyo Institute of Technology, Tokyo 152-8551, Japan}

\begin{abstract}
The covariant St\"{u}ckelberg formalism is applied to bi-gravity in
order to revisit the issue of absence of the Boulware-Deser (BD)
ghost. We first confirm that the leading order action in the decoupling
limit for helicity-2 modes of metrics and helicity-0 mode of
St\"{u}ckelberg perturbations does not lead to higher
time derivative in equations of motion, which suggests the absence of
the BD ghost. By extending this method, we reconfirm that the BD ghost
does not appear for arbitrary order of the perturbations at the
decoupling limit in bi-gravity.
\end{abstract}

\maketitle
\section{Introduction}

One possible way of modification of general relativity is to introduce
extra degrees of freedom to gravitation. In order for this kind of
extension to be viable, these extra degrees of freedom must not
destabilize the system, especially, must not be ghosts.

To give a mass to graviton is a potent way to add degrees of freedom to
gravitation. On linear metric perturbations around the flat fiducial
metric, Fierz and Pauli(FP) establish the theory which has five
gravitational degrees of freedom without ghosts~\cite{Fierz:1939ix}.
The theory consisting of the non-linear Einstein-Hilbert kinetic term
and the FP mass potential, however, excites six degrees of freedom and
the additional one is the Boulware-Deser(BD)
ghost~\cite{Boulware:1973my}. In the St\"{u}ckelberg formalism with
taking decoupling limit~\cite{ArkaniHamed:2002sp}, the BD ghost
instability can be regarded as the Ostrogradsky
instability~\cite{Woodard:2015zca} associated with the dangerous higher
time derivative of the helicity-0 mode of the St\"{u}ckelberg fields.
de Rham, Gabadadze and Tolley(dRGT) construct the mass potential where
the self interactions of the helicity-0 mode take the Galileon
form~\cite{Nicolis:2008in} and hence there is no Ostrogradsky
instability at least in decoupling
limit~\cite{deRham:2010ik,deRham:2010kj,Ondo:2013wka}. It is finally proven that dRGT
theory is free from the BD ghost even without taking decoupling
limit~\cite{Hassan:2011hr,Hassan:2011ea,Kugo:2014hja}.

The theory of massive gravity is first constructed with the flat
fiducial metric. dRGT massive gravity with the flat fiducial metric can
be extend to that with a general fiducial metric and this theory is also
proven to be BD ghost free~\cite{Hassan:2011ea,Hassan:2011tf}.  Since
the full theory is BD ghost free, the equations of motion of the
helicity-0 mode in the decoupling limit should not include higher time derivatives.
In Ref.~\cite{Gao:2014ula}, two of the present authors and their collaborators confirm this fact directly by formulating the covariant St\"{u}ckelberg analysis for general fiducial metric.
There, non-trivial couplings between the curvature of fiducial metric and
helicity-0 mode of St\"{u}ckelberg fields appear. Such coupling terms,
however, do not produce higher time derivatives in the equations of
motion because the fiducial metric is non-dynamical in massive gravity.

Massive gravity was further extended to bi-gravity, in which the
Einstein-Hilbert kinetic term of a fiducial metric is added to dRGT
massive gravity, and this bi-gravity theory is also shown to be BD ghost
free~\cite{Hassan:2011ea,Hassan:2011zd}. Then, the equations of motion
in decoupling limit should not include higher time derivatives. On the
other hand, by the analogy with the covariant St\"{u}ckelberg analysis
of massive gravity, there should be non-trivial derivative couplings
between the fiducial metric and the helicity-0 mode. The purpose of this
short note is to clarify why such higher derivative interactions do not
cause the Ostrogradsky instability.

In the next section, we will give a brief review of the
covariant St\"{u}ckelberg analysis of dRGT massive gravity with a
general fiducial metric established in Ref.~\cite{Gao:2014ula}. Then, in
Sec. \ref{sec:bigravity}, we will extend the results of dRGT theory to
bi-gravity. Final section is devoted to conclusions.

\section{Covariant St\"{u}ckelberg formalism of massive gravity}
\label{sec:massive}

We begin with a brief review of the covariant St\"{u}ckelberg formalism
of massive gravity with a general fiducial metric established in
Ref.~\cite{Gao:2014ula}. The action of dRGT massive gravity with a
general fiducial metric $\bar{g}_{\mu\nu}$ is given as
follows~\cite{deRham:2010kj,Hassan:2011tf,Hassan:2011vm}:
\begin{eqnarray}
 S[g,\bar{g}] = \frac{M_{\rm pl}^2}{2}\int d^4 x \sqrt{-g} \left(R[g] +2 m^2
					     \sum_{i=0}^{4} \beta_i
					     e_i(\gamma[g,\bar{g}])\right),
\label{Smg}
\end{eqnarray}
where
\begin{eqnarray}
 \gamma[g,\bar{g}]^{\mu}{}_{\nu} = \sqrt{g^{-1}\bar{g}}^{\mu}{}_{\nu},
\end{eqnarray}
and
\begin{eqnarray}
 e_{i}[\gamma] = \gamma^{\mu_{1}}{}_{[\mu_{1}}\cdots \gamma^{\mu_{n}}{}_{\mu_{n}]}.
\end{eqnarray}
In this note, we focus on the following parameters used in the original
dRGT theory~\cite{deRham:2010kj}: \footnote{In the original dRGT theory,
the action is written in terms of $\alpha$ parameters as
\begin{eqnarray*}
 S[g,\bar{g}]
&=&\frac{M_{\rm pl}^2}{2}\int d^4 x \left[~\sqrt{-g} R[g] + \sqrt{-g} m^2
\sum_{i=2}^{4} \left({\rm i}! \right) \alpha_i e_i({\cal K})~\right],
\end{eqnarray*}
with $ {\cal K}^{\mu}{}_{\nu} = \delta^{\mu}{}_{\nu}
-\gamma^{\mu}{}_{\nu} $ and $\alpha_2 = 1$.}
\begin{eqnarray}
 \beta_0 &=& 6 + 12 \alpha_3 +12 \alpha_4,\\
 \beta_1 &=& -3-9\alpha_3 -12 \alpha_4,\\
 \beta_2 &=& 1 +6 \alpha_3 +12 \alpha_4,\\
 \beta_3 &=& -3 \alpha_3 -12 \alpha_4,\\
 \beta_4 &=&12 \alpha_4.
\end{eqnarray}
Due to the presence of the mass term, the action (\ref{Smg}) does not
possess the gauge symmetry on diffeomorphism.  However, we can rewrite
the action~(\ref{Smg}) as gauge invariant one by introducing
St\"{u}ckelberg fields $\phi^{a}$ with $a =0,1,2,3$, and by replacing
the original fiducial metric with covariantized one:
\begin{eqnarray}
\bar{g}_{\mu\nu} \rightarrow  f^{\phi}_{\mu\nu}(x) = \frac{\partial \phi^a(x)}{\partial x^{\mu}} \frac{\partial \phi^b(x)}{\partial x^{\nu}} \bar{g}_{ab}(\phi).
\end{eqnarray}
The resultant action
\begin{eqnarray}
 S = \frac{M_{\rm pl}^2}{2}\int d^4 x \sqrt{-g} \left(R[g] +2 m^2 \sum_{i=0}^{4} \beta_i e_i(\gamma[g,f^{\phi}])\right)
\end{eqnarray}
is invariant under the following gauge transformation,
\begin{eqnarray}
 g_{\mu\nu}(x) \rightarrow g'_{\mu\nu}(x) &=& g_{\alpha\beta}(y(x))\frac{\partial y^{\alpha}(x)}{\partial x^{\mu}}\frac{\partial y^{\beta}(x)}{\partial x^{\nu}}, \\
 \phi^a (x) \rightarrow \phi'{}^a(x) &=& \phi^a(y(x)).
\end{eqnarray}
Here, the original action~(\ref{Smg}) can be recovered by fixing the
gauge $\phi^a (x) = x^a $, which is called unitary gauge.

We consider perturbations of St\"{u}ckelberg fields around the unitary
gauge. When the fiducial metric is flat, $\bar{g}_{\mu\nu} =
\eta_{\mu\nu}$, the perturbations $\tilde{\pi}^a$ simply defined by
$\phi^a(x) = x^a -\tilde{\pi}^a $ are well behaved because
$\tilde{\pi}^a$ are covariant under the Lorentz transformation. In the
non-linear theory consisting of the Einstein-Hilbert action and the
linear FP mass potential, the BD ghost instability appears as the
Ostrogradsky instability in the form of higher time derivative in the
equation of motion of helicity-0 mode of $\tilde{\pi}^a$. Therefore,
absence of such higher time derivatives is a necessary condition
for being 
free from the BD ghost.

In the case of a general fiducial metric, since $\tilde{\pi}^a$ is not a covariant vector in curved spacetime, we need
to refine the definition of St\"{u}ckelberg perturbations.\footnote{ The
decoupling limit analysis based on the embedding method is investigated
in~\cite{deRham:2012kf} for de Sitter fiducial metric. This method is
equivalent to our Riemann normal coordinate approach as proven
in~\cite{Gao:2015xwa}.  Another decoupling limit analysis based on the
vielbein formalism is investigated in \cite{Fasiello:2013woa}.  } We
have defined $\pi^a$ as a coordinate value of a point $\phi^a$ in the
field space with the Riemann normal coordinate on the fiducial metric
$\bar{g}_{ab}$ \cite{Gao:2014ula} , that is,
\begin{eqnarray}
 \phi^a = x^a - \pi^a -\frac{1}{2}\bar{\Gamma}^{a}{}_{bc}\pi^{b} \pi^c+\frac{1}{6}\left(\partial_{b}\bar{\Gamma}^{a}{}_{cd}-2 \bar{\Gamma}^{a}{}_{be} \bar{\Gamma}^{e}{}_{cd}\right)\pi^b \pi^c \pi^d + {\cal O}(\epsilon^4),
\end{eqnarray}
where $\epsilon$ represents the order of perturbations. From the
definition of the Riemann normal coordinate, $\pi^a$ is a covariant
vector. By this definition, the covariantized fiducial metric is
expanded in a covariant way:
\begin{eqnarray}
 f_{\mu\nu}^{\phi} &=& \bar{g}_{\mu\nu} - 2 \bar{\nabla}_{(\mu}\pi_{\nu)}+\bar{\nabla}_{\mu}\pi_{\rho}\bar{\nabla}_{\nu}\pi^{\rho} - \bar{R}_{\mu\rho\nu\sigma}\pi^\rho \pi^{\sigma}  \notag\\
 &&
  +\frac{1}{3}\bar{\nabla}_\lambda\bar{R}_{\mu\rho\nu\sigma}\pi^\lambda \pi^\rho \pi^\sigma
+\frac{2}{3}\bar{R}_{\mu\rho\lambda\sigma}\bar{\nabla}_\nu \pi^\lambda \pi^\rho \pi^\sigma
+
\frac{2}{3}\bar{R}_{\nu\rho\lambda\sigma}\bar{\nabla}_\mu \pi^\lambda \pi^\rho \pi^\sigma+ {\cal O}(\epsilon^4),\label{fphi}
\end{eqnarray}
where $\pi_{\mu} = \bar{g}_{\mu\nu}\pi^\nu$.  Another derivation of
Eq.~(\ref{fphi}) is investigated in~\cite{Gao:2015xwa}.  Hereafter, we
concentrate only on the helicity-0 mode $\hat{\pi}$ defined by
\begin{eqnarray}
 \pi_{\mu} := \frac{\bar{\nabla}_{\mu} \hat{\pi}}{m^2 M_{\rm pl}}
\end{eqnarray}
to focus on the presence/absence of the BD ghost. We also consider the
metric perturbations around the fiducial metric:
\begin{eqnarray}
 g_{\mu\nu} &=& \bar{g}_{\mu\nu} +  \frac{\hat{h}_{\mu\nu}}{M_{\mathrm{pl}}}.
\end{eqnarray}
Here $\hat{\pi}$ and $\hat{h}_{\mu\nu}$ have canonically
normalized dimensions.

After taking the extended $\Lambda_3$ decoupling
limit~\cite{Gao:2014ula},
\begin{eqnarray}
 M_{\mathrm{pl}} \rightarrow \infty,\
 m \rightarrow 0,\
 \Lambda_3 := (M_{\mathrm{pl}} m^2 )^{1/3} \rightarrow \text{finite},\
 \frac{\bar{R}_{\mu\nu\rho\sigma}}{m^2} \rightarrow \text{finite},\label{decouplemg}
\end{eqnarray}
the action reduces to
\begin{eqnarray}
 S &=&\int d^4 x \sqrt{-\bar{g}}\left(-\frac{1}{4}\hat{h}_{\mu\nu} \bar{{\cal E}}^{\mu\nu\rho\sigma} \hat{h}_{\rho\sigma} + {\cal L}^{mass}[\bar{g},\hat{h},\hat{\pi}]\right).
\end{eqnarray}
Here, the first term represents the contribution from the
Einstein-Hilbert action, where the operators $\bar{{\cal
E}}^{\mu\nu\rho\sigma}$ are given by
\begin{eqnarray}
 \bar{\mathcal{E}}^{\mu\nu\rho\sigma}\hat{h}_{\rho\sigma} & = & -\frac{1}{2}\bar{\square}\hat{h}^{\mu\nu}-\frac{1}{2}\bar{\nabla}^{\mu}\bar{\nabla}^{\nu}\hat{h}+\frac{1}{2}\bar{g}^{\mu\nu}\left(\bar{\square}\hat{h}-\bar{\nabla}_{\rho}\bar{\nabla}_{\sigma}\hat{h}^{\rho\sigma}\right)+\bar{\nabla}_{\rho}\bar{\nabla}^{(\mu}\hat{h}^{\nu)\rho}.\label{L_EH_linear}
\end{eqnarray}
On the other hand, ${\cal L}^{mass}$ represents the contribution from
the dRGT mass potential and is concretely given as
\begin{eqnarray}
{\cal L}^{mass}[\bar{g},\hat{h},\hat{\pi}] &=&   \frac{1}{2}\hat{h}^{\mu\nu}X^{(1)}_{\mu\nu}(\hat{\pi}) +  \frac{1}{2} \frac{\bar{R}^{\mu\nu}}{m^2}\bar{\nabla}_{\mu}\hat{\pi}\bar{\nabla}_{\nu}\hat{\pi}
  + \frac{1+ 3 \alpha_3}{4 \Lambda^3_3}\hat{h}^{\mu\nu}X^{(2)}_{\mu\nu}(\hat{\pi}) +  \frac{1}{2 \Lambda^3_3} {\cal A}_{\mu\nu\rho\sigma}\bar{\nabla}^\mu \hat{\pi}\bar{\nabla}^\nu \hat{\pi}\bar{\nabla}^\rho\bar{\nabla}^\sigma \hat{\pi}\notag\\
 && +  \frac{\alpha_3+4\alpha_4}{4 \Lambda^6_3}\hat{h}^{\mu\nu}X^{(3)}_{\mu\nu}(\hat{\pi})\notag\\
 &&+  \frac{1}{2 \Lambda^6_3}\left({\cal B}_{\mu\nu\rho\sigma\rho'\sigma'}\bar{\nabla}^{\rho'}\bar{\nabla}^{\sigma'}\hat{\pi}-\frac{1}{3}{\cal C}_{\lambda\mu\nu\rho\sigma}\bar{\nabla}^{\lambda}\hat{\pi}\right)\bar{\nabla}^\mu \hat{\pi}\bar{\nabla}^\nu \hat{\pi}\bar{\nabla}^\rho\bar{\nabla}^\sigma \hat{\pi} + O(\epsilon^5),\label{Lmassgbar}
\end{eqnarray}
where $X_{\mu\nu}^{(i)}$ are defined by
\begin{eqnarray}
X_{\mu\nu}^{(1)}\left(\hat{\pi}\right) & = & \bar{g}_{\mu\nu}\bar{\square}\hat{\pi}-\bar{\nabla}_{\mu}\bar{\nabla}_{\nu}\hat{\pi},\label{X1}\\
X_{\mu\nu}^{(2)}\left(\hat{\pi}\right) & = & \bar{g}_{\mu\nu}\left(\left(\bar{\square}\hat{\pi}\right)^{2}-\bar{\nabla}_{\rho}\bar{\nabla}_{\sigma}\hat{\pi}\bar{\nabla}^{\rho}\bar{\nabla}^{\sigma}\hat{\pi}\right)+2\left(\bar{\nabla}_{\mu}\bar{\nabla}_{\rho}\hat{\pi}\bar{\nabla}^{\rho}\bar{\nabla}_{\nu}\hat{\pi}-\bar{\square}\hat{\pi}\bar{\nabla}_{\mu}\bar{\nabla}_{\nu}\hat{\pi}\right),\label{X2}\\
X_{\mu\nu}^{(3)}\left(\hat{\pi}\right) & = & \bar{g}_{\mu\nu}\left(\left(\bar{\square}\hat{\pi}\right)^{3}-3\bar{\square}\hat{\pi}\bar{\nabla}_{\rho}\bar{\nabla}_{\sigma}\hat{\pi}\bar{\nabla}^{\rho}\bar{\nabla}^{\sigma}\hat{\pi}+2\bar{\nabla}^{\rho}\bar{\nabla}_{\sigma}\hat{\pi}\bar{\nabla}^{\sigma}\bar{\nabla}_{\lambda}\hat{\pi}\bar{\nabla}^{\lambda}\bar{\nabla}_{\rho}\hat{\pi}\right)\nonumber \\
 &  & +3\bar{\nabla}_{\mu}\bar{\nabla}_{\nu}\hat{\pi}\left(\bar{\nabla}_{\rho}\bar{\nabla}_{\sigma}\hat{\pi}\bar{\nabla}^{\rho}\bar{\nabla}^{\sigma}\hat{\pi}-\left(\bar{\square}\hat{\pi}\right)^{2}\right)
  \notag\\
 && +6\bar{\nabla}^{\rho}\bar{\nabla}_{\mu}\hat{\pi}\left(\bar{\nabla}_{\nu}\bar{\nabla}_{\rho}\hat{\pi}\bar{\square}\hat{\pi}-\bar{\nabla}_{\nu}\bar{\nabla}^{\sigma}\hat{\pi}\bar{\nabla}_{\rho}\bar{\nabla}_{\sigma}\hat{\pi}\right),\label{X3}
\end{eqnarray}
and ${\cal A}, {\cal B}$ and ${\cal C}$, which consist of the curvature of the fiducial metric, are defined by
	\begin{eqnarray}
	\mathcal{A}_{\mu\nu\rho\sigma} & = & \frac{1}{m^{2}}\left[\left(1+2\alpha_{3}\right)\left(\bar{R}_{\mu\nu}\bar{g}_{\rho\sigma}+\bar{R}_{\rho(\mu\nu)\sigma}\right)-\alpha_{3}\left(\bar{g}_{\rho(\mu}\bar{R}_{\nu)\sigma}+\bar{g}_{\sigma(\mu}\bar{R}_{\nu)\rho}\right)\right],\label{A_def}\\
	\mathcal{B}_{\mu\nu\rho\sigma\rho'\sigma'} & = & \frac{1}{m^{2}}\bigg[\frac{3}{2}\left(\alpha_{3}+2\alpha_{4}\right)\bar{R}_{\mu\nu}\left(2\bar{g}_{\rho[\sigma}\bar{g}_{\sigma']\rho'}\right)+12\alpha_{4}\bar{R}_{\mu[\rho}\bar{g}_{\rho'][\sigma}\bar{g}_{\sigma']\nu}\nonumber \\
	 &  & \qquad\qquad -\frac{1}{3}\left(1+9\alpha_{3}+18\alpha_{4}\right)\left(\bar{R}_{\mu\rho\nu[\sigma}\bar{g}_{\sigma']\rho'}-\bar{R}_{\mu\rho'\nu[\sigma}\bar{g}_{\sigma']\rho}\right)
	  \notag\\
	 && \qquad \qquad -6\alpha_{4}\bar{g}_{\mu[\rho}\bar{R}_{\rho']\nu\sigma\sigma'}\bigg], \label{B_def}\\
	\mathcal{C}_{\lambda\mu\nu\rho\sigma} & = & \frac{1}{m^{2}}\left[\bar{g}_{\rho\sigma}\bar{\nabla}_{(\lambda}\bar{R}_{\mu\nu)}+\frac{1}{3}\left(\bar{\nabla}_{\lambda}\bar{R}_{\mu(\rho\sigma)\nu}+\bar{\nabla}_{\mu}\bar{R}_{\lambda(\rho\sigma)\nu}+\bar{\nabla}_{\nu}\bar{R}_{\lambda(\rho\sigma)\mu}\right)\right].\label{C_def}
\end{eqnarray}
It should be noted that we implicitly assume $g_{\mu\nu}
= \bar{g}_{\mu\nu}$ and $ \phi^\mu = x^{\mu}$ is a solution of the
background equations of motion in order for the linear order action to
vanish. If this is not a vacuum solution, we should assume some matter
fields to guarantee the absence of tadpole contributions.

The mixing terms $\hat{h}^{\mu\nu}X^{(1)}_{\mu\nu}$ and
$\hat{h}^{\mu\nu}X^{(2)}_{\mu\nu}$ can be diagonalized by the field
redefinition,\footnote{The field redefinition (\ref{redef}) contains derivatives, so one may wonder if it changes the number of physical degrees of freedom. However, as we discuss in Sec. \ref{sec:bigravity}, the degrees of freedom in the perturbation theory can be determined only from the quadratic action. Since derivatives in (\ref{redef}) appear at the nonlinear level, it turns out that the number of degrees of freedom does not change at any order in the perturbation theory.}
\begin{equation}
\hat{h}_{\mu\nu} \rightarrow \hat{h}_{\mu\nu} + \hat{\pi} \bar{g}_{\mu\nu} - \frac{1+3\alpha_3}{\Lambda_3^3} \bar{\nabla}_{\mu}\hat{\pi}\bar{\nabla}_{\nu}\hat{\pi}.\label{redef}
\end{equation}
The resultant action becomes
\begin{eqnarray}
 S = \int d^4 x\sqrt{-\bar{g}} \left(-\frac{1}{4}\hat{h}_{\mu\nu}\bar{\mathcal{E}}^{\mu\nu\rho\sigma}\hat{h}_{\rho\sigma}+{\cal L}^{\text{G}} + {\cal L}^{\text{C}} +\frac{1}{4\Lambda_{3}^{6}}\left(\alpha_{3}+4\alpha_{4}\right)\hat{h}^{\mu\nu}X_{\mu\nu}^{(3)}\left(\hat{\pi}\right) +{\cal O}(\epsilon^5)\right),
\end{eqnarray}
where
\begin{eqnarray}
 {\cal L}^{\text{G}}
  &=&
  -\frac{3}{4}\bar{g}_{\mu\nu}\bar{\nabla}^{\mu}\hat{\pi}\bar{\nabla}^{\nu}\hat{\pi}
  -\frac{3\left(1+3\alpha_{3}\right)}{4\Lambda_{3}^{3}}\left(\bar{\nabla}\hat{\pi}\right)^{2}\bar{\square}\hat{\pi}
  \notag\\
 &&-\frac{1+8\alpha_{3}+9\alpha_{3}^{2}+8\alpha_{4}}{4\Lambda_{3}^{6}}\left(\bar{\nabla}\hat{\pi}\right)^{2}\left(\left(\bar{\square}\hat{\pi}\right)^{2}-\bar{\nabla}_{\rho}\bar{\nabla}_{\sigma}\hat{\pi}\bar{\nabla}^{\rho}\bar{\nabla}^{\sigma}\hat{\pi}\right),\\
 \mathcal{L}^{\text{C}}&=&\frac{1}{2}\frac{\bar{R}_{\mu\nu}}{m^{2}}\bar{\nabla}^{\mu}\hat{\pi}\bar{\nabla}^{\nu}\hat{\pi}
  +\frac{1}{2\Lambda_{3}^{3}}\mathcal{A}_{\mu\nu\rho\sigma}\bar{\nabla}^{\mu}\hat{\pi}\bar{\nabla}^{\nu}\hat{\pi}\bar{\nabla}^{\rho}\bar{\nabla}^{\sigma}\hat{\pi}
  \notag\\&&
  +\frac{1}{2\Lambda_{3}^{6}}\left(\mathcal{B}_{\mu\nu\rho\sigma\rho'\sigma'}\bar{\nabla}^{\rho'}\bar{\nabla}^{\sigma'}\hat{\pi}-\frac{1}{3}\mathcal{C}_{\lambda\mu\nu\rho\sigma}\bar{\nabla}^{\lambda}\hat{\pi}\right)\bar{\nabla}^{\mu}\hat{\pi}\bar{\nabla}^{\nu}\hat{\pi}\bar{\nabla}^{\rho}\bar{\nabla}^{\sigma}\hat{\pi}.\label{L_unmix_4}
\end{eqnarray}
Since the derivative operator $\bar{\nabla}_\mu$ is commutative in
decoupling limit, the covariant Galileon term ${\cal L}^{\text{G}}$ does
not lead to higher time derivative terms in the equation of motion of
helicity-0 mode of $\tilde{\pi}^a$. In addition, as found in
Ref.~\cite{Gao:2014ula}, the curvature term ${\cal L}^{\text{C}}$ does
not produce higher derivative terms either. Thus, no Ostrogradsky
instability appears at least up to fourth order of perturbations.

\section{Covariant St\"{u}ckelberg formalism of bi-gravity}
\label{sec:bigravity}

The purpose of this note is to investigate the Ostrogradsky instability
of helicity-0 mode of St\"{u}ckelberg fields in bi-gravity by extending
the discussions given in the previous section. The effect of the
interactions between the curvature of the fiducial metric and
$\hat{\pi}$ is not clear because the fiducial metric itself is dynamical
in bi-gravity. While the covariant Galileon term ${\cal L}^{\text{G}}$
does not lead to higher time derivative terms in the equation of motion
of the fiducial metric as well as that of helicity-0 mode of
$\tilde{\pi}^a$, the curvature term ${\cal L}^{\text{C}}$ naively
generates higher time derivative terms in the equations of motion, which
might be dangerous.
We show, however, that there are no ghosts at the linear perturbation level, so that bi-gravity is free from the Ostrogradsky instability at any order in the perturbation theory.

Let us consider the theory of bi-gravity~\cite{Hassan:2011zd}, in which
the action is given by
\begin{eqnarray}
 S = \frac{M_{\rm pl}^2}{2}\int d^4 x \sqrt{-g} \left(R[g] +2 m^2 \sum_{i=0}^{4} \beta_i e_i(\gamma[g,f])\right) +  \frac{\kappa^2 M_{\rm pl}^2}{2}\int d^4 x \sqrt{-f}R[f].\label{Sbg}
\end{eqnarray}
First we introduce St\"{u}ckelberg fields by analogy with massive gravity. The action of bi-gravity is invariant under the gauge transformation
\begin{eqnarray}
 g_{\mu\nu}(x) \rightarrow g'_{\mu\nu}(x) &=& g_{\alpha\beta}(y(x))\frac{\partial y^{\alpha}(x)}{\partial x^{\mu}}\frac{\partial y^{\beta}(x)}{\partial x^{\nu}}, \label{eq:gauge1}\\
 f_{\mu\nu}(x) \rightarrow f'_{\mu\nu}(x) &=& f_{\alpha\beta}(y(x))\frac{\partial y^{\alpha}(x)}{\partial x^{\mu}}\frac{\partial y^{\beta}(x)}{\partial x^{\nu}}.\label{eq:gauge2}
\end{eqnarray}
St\"{u}ckelberg formalism of bi-gravity is obtained by the following
replacement,
\begin{eqnarray}
\gamma[g,f]\to\gamma[g,f^\phi]
\quad {\rm with}\quad
  f^{\phi}_{\mu\nu}(x) = f_{\rho\sigma}(\phi(x))\partial_{\mu} \phi^\rho \partial_{\nu} \phi^\sigma.
\end{eqnarray}
The resultant action,
\begin{eqnarray}
 S = \frac{M_{\rm pl}^2}{2}\int d^4 x \sqrt{-g} \left(R[g] +2m^2 \sum_{i=0}^{4} \beta_i e_i(\gamma[g,f^\phi])\right)+  \frac{\kappa^2 M_{\rm pl}^2}{2}\int d^4 x \sqrt{-f}R[f],\label{SbgSt}
\end{eqnarray}
is invariant under the following two gauge symmetries. The first one is
given by
\begin{eqnarray}
  g_{\mu\nu}(x) \rightarrow g'_{\mu\nu}(x) &=& g_{\alpha\beta}(y(x))\frac{\partial y^{\alpha}(x)}{\partial x^{\mu}}\frac{\partial y^{\beta}(x)}{\partial x^{\nu}},\\
  f_{\mu\nu}(x) \rightarrow f'_{\mu\nu}(x) &=& f_{\mu\nu}(x),\\
  \phi^\mu(x) \rightarrow \phi'{}^{\mu}(x) &=& \phi^{\mu}(y(x)),
\end{eqnarray}
which leads to
\begin{eqnarray}
  f^\phi_{\mu\nu}(x) \rightarrow f^{\phi\,\prime}_{\mu\nu}(x) = f^\phi_{\alpha\beta}(y(x))\frac{\partial y^{\alpha}(x)}{\partial x^{\mu}}\frac{\partial y^{\beta}(x)}{\partial x^{\nu}}.
\end{eqnarray}
The second one is given by
\begin{eqnarray}
  g_{\mu\nu}(x) \rightarrow g'_{\mu\nu}(x) &=& g_{\mu\nu}(x),\\
  f_{\mu\nu}(x) \rightarrow f'_{\mu\nu}(x) &=& f_{\alpha\beta}(z(x))\frac{\partial z^{\alpha}(x)}{\partial x^{\mu}}\frac{\partial z^{\beta}(x)}{\partial x^{\nu}},\\
  \phi^\mu(x) \rightarrow \phi'{}^{\mu}(x) &=& (z^{-1})^{\mu}(\phi(x)).
\end{eqnarray}
which leads to
\begin{eqnarray}
  f^\phi_{\mu\nu}(x) \rightarrow f^{\phi\,\prime}_{\mu\nu}(x) = f^\phi_{\mu\nu}(x).
\end{eqnarray}
The original action~(\ref{SbgSt}) can be recovered by fixing the unitary
gauge $\phi^\mu = x^\mu$ and a combination of these gauge
transformations with $y^{\alpha}(x)=z^{\alpha}(x)$ reproduces the gauge
transformation with Eqs.~(\ref{eq:gauge1}) and (\ref{eq:gauge2}).

In order to take the decoupling limit of the action in bi-gravity, we
consider metric perturbations around $g_{\mu\nu} = \bar{g}_{\mu\nu}$ and
$f_{\mu\nu} = \bar{g}_{\mu\nu}$,
\begin{eqnarray}
  g_{\mu\nu} &=& \bar{g}_{\mu\nu} + \frac{\hat{l}_{\mu\nu}}{M_{\rm pl}\kappa} +  \frac{\hat{h}_{\mu\nu}}{M_{\mathrm{pl}}}\\
&=&   f_{\mu\nu} +  \frac{\hat{h}_{\mu\nu}}{M_{\mathrm{pl}}},\\
  f_{\mu\nu} &=& \bar{g}_{\mu\nu} + \frac{\hat{l}_{\mu\nu}}{M_{\rm pl}\kappa}
\end{eqnarray}
and covariant helicity-0 St\"{u}ckelberg perturbations around unitary gauge,
\begin{eqnarray}
  \phi^\mu &=& x^\mu - \pi^{\mu} -\frac{1}{2}{\Gamma^{(f)}{}^{\mu}_{\nu\rho}}\pi^\nu \pi^\rho 
+\frac{1}{6}\left(\partial_{b}\Gamma^{(f)a}{}_{cd}-2 \Gamma^{(f)a}{}_{be} \Gamma^{(f)e}{}_{cd}\right)\pi^b \pi^c \pi^d + {\cal O}(\epsilon^4),
\end{eqnarray}
with
\begin{eqnarray}
  \pi^{\mu} = \frac{f^{\mu\nu}\nabla^{(f)}_{\nu} \hat{\pi}}{m^2 M_{\rm pl}}.
\end{eqnarray}
The decoupling limit~(\ref{decouplemg}) of two Einstein-Hilbert actions reduces to
\begin{eqnarray}
  &&  \frac{M_{\rm pl}^2}{2}\int d^4x \sqrt{-g} R[g] +    \frac{\kappa^2 M_{\rm pl}^2}{2}\int d^4x \sqrt{-f} R[f]
   \notag\\
   &=&
    \int d^4 x \sqrt{-\bar{g}} \Big(-\frac{1}{4}(\hat{h}_{\mu\nu} + {\kappa}^{-1}\hat{l}_{\mu\nu})\bar{{\cal E}}^{\mu\nu\rho\sigma}(\hat{h}_{\rho\sigma}+{\kappa}^{-1}\hat{l}_{\rho\sigma}) -\frac{1}{4}\hat{l}_{\mu\nu}\bar{{\cal E}}^{\mu\nu\rho\sigma}\hat{l}_{\rho\sigma} \Big).
\end{eqnarray}
Since Riemann tensor of $f_{\mu\nu}$ is related with that of $\bar{g}_{\mu\nu}$ as
\begin{eqnarray}
 \frac{R^{(f)}_{\mu\nu\rho\sigma}}{m^2} = \frac{\bar{R}_{\mu\nu\rho\sigma}}{m^2} + \frac{1}{\Lambda^3_3}\frac{1}{2\kappa}\left(
\bar{\nabla}_{\sigma}\bar{\nabla}_{\mu}\hat{l}_{\nu\rho}-\bar{\nabla}_{\rho}\bar{\nabla}_{\mu}\hat{l}_{\nu\sigma}-\bar{\nabla}_{\sigma}\bar{\nabla}_{\nu}\hat{l}_{\mu\rho}+\bar{\nabla}_{\rho}\bar{\nabla}_{\nu}\hat{l}_{\mu\sigma}
 \right) + {\cal O}\left(\frac{1}{\Lambda^3_3M_{\mathrm{pl}}}\right),
\end{eqnarray}
our decoupling limit~(\ref{decouplemg}) is equivalent to
 \begin{eqnarray}
   M_{\mathrm{pl}} \rightarrow \infty,\
 m \rightarrow 0,\
 \Lambda_3 := (M_{\mathrm{pl}} m^2 )^{1/3} \rightarrow \text{finite},\
 \frac{R^{(f)}_{\mu\nu\rho\sigma}}{m^2} \rightarrow \text{finite}.
 \end{eqnarray}

Then, the decoupling limit of interaction terms is given by replacing
$\bar{g}_{\mu\nu}$ in~(\ref{Lmassgbar}) with $f_{\mu\nu}$ as
\begin{eqnarray}
 \frac{M_{\rm pl}^2}{2}\int d^4 x \sqrt{-g} \left(2 m^2 \sum_{i=0}^{4} \beta_i e_i(\gamma[g,f^\phi])\right) =  \int d^4 x \sqrt{-f} {\cal L}^{mass}[f,\hat{h},\hat{\pi}].
\end{eqnarray}
Obviously, the interaction terms have non-trivial higher derivative couplings, for example,   
\begin{eqnarray}
&& \int d^4 x \sqrt{-f} \frac{R^{(f)}{}^{\mu\nu}}{m^2} \nabla^{(f)}_{\mu} \hat{\pi} \nabla^{(f)}_{\nu} \hat{\pi} \notag\\
 &=& \int d^4 x \sqrt{-\bar{g}}\left[ \frac{\bar{R}^{\mu\nu}}{m^2} \bar{\nabla}_{\mu} \pi \bar{\nabla}_{\nu} \pi + \frac{1}{\Lambda^3_3} \frac{1}{2\kappa}\left(
2\bar{\nabla}_{\sigma}\bar{\nabla}_{\mu}\hat{l}_{\nu}{}^{\mu}-\bar{\Box}\hat{l}_{\nu\sigma} -\bar{\nabla}_{\sigma}\bar{\nabla}_{\nu}\hat{l}
 \right)\bar{\nabla}^{\nu} \hat{\pi} \bar{\nabla}^{\sigma} \hat{\pi}\right].\label{ddldpdp}
\end{eqnarray}
The variation of (\ref{ddldpdp}) produces the terms with
third order time derivatives. At a
glance, this result seems to contradict the absence of BD ghost in the
full non-perturbative theory.  However, the key observation here is that
such higher derivative interactions do not appear at the leading order
of ${\cal L}^{mass}$, because
\begin{eqnarray}
  \int d^4 x \sqrt{-f}{\cal L}^{mass}[f,\hat{h},\hat{\pi}]
   = \int d^4 x \sqrt{-\bar{g}}{\cal L}^{mass}[\bar{g},\hat{h},\hat{\pi}] + {\cal O}(\epsilon^3).
\end{eqnarray}
Then the decoupling limit of the total action can be written as
\begin{eqnarray}
  S &=&  \int d^4 x \sqrt{-\bar{g}} \Big(-\frac{1}{4}(\hat{h}_{\mu\nu} + {\kappa}^{-1}\hat{l}_{\mu\nu})\bar{{\cal E}}^{\mu\nu\rho\sigma}(\hat{h}_{\rho\sigma}+{\kappa}^{-1}\hat{l}_{\rho\sigma}) -\frac{1}{4}\hat{l}_{\mu\nu}\bar{{\cal E}}^{\mu\nu\rho\sigma}\hat{l}_{\rho\sigma}\notag\\
  &&\qquad+ {\cal L}^{mass}[\bar{g},\hat{h},\hat{\pi}] + {\cal O}(\epsilon^3)  \Big).
\end{eqnarray}
By introducing
\begin{eqnarray}
  L_{\mu\nu} &=& \frac{1}{\sqrt{1+\kappa^2}}\left(\hat{h}_{\mu\nu}+\left(\kappa+\frac{1}{\kappa}\right)\hat{l}_{\mu\nu}\right),
\end{eqnarray}
the total action reduces to at leading order ${\cal O}(\epsilon^2)$
\begin{eqnarray}
 S^{(2)}[\hat{h},\hat{l},\hat{\pi}] &=& \int d^4 x \sqrt{-\bar{g}} \Big(-\frac{\kappa^2}{1+\kappa^2}\frac{1}{4}\hat{h}_{\mu\nu}\bar{{\cal E}}^{\mu\nu\rho\sigma}\hat{h}_{\rho\sigma} + \frac{1}{2}\hat{h}^{\mu\nu}X^{(1)}_{\mu\nu}(\hat{\pi}) + \frac{1}{2}\frac{\bar{R}^{\mu\nu}}{m^2}\bar{\nabla}_{\mu}\hat{\pi}\bar{\nabla}_{\nu}\hat{\pi} - \frac{1}{4}L_{\mu\nu}\bar{{\cal E}}^{\mu\nu\rho\sigma}L_{\rho\sigma}\Big).\notag\\\label{S2}
\end{eqnarray}
It is manifest that $\hat{h}$ and $\hat{\pi}$ are decoupled from
$L_{\mu\nu}$ at leading order. The first three terms in Eq.~(\ref{S2})
coincide with the decoupling limit of the action in massive gravity when
we expand $g_{\mu\nu} = \bar{g}_{\mu\nu} +
\hat{h}_{\mu\nu}/M_{\mathrm{pl}}$ and the Einstein-Hilbert term has
additional coefficient $\kappa^2/(1+\kappa^2)$. The last term coincides
with the decoupling limit of the Einstein-Hilbert action. Thus, linear
perturbations are free from the Ostrogradsky instability even in
bi-gravity. This is one of the main conclusions of this note. It should be
noticed that $\bar{R}_{\mu\nu}$ in the third term comes from a
non-dynamical field $\bar{g}_{\mu\nu}$. The dynamical degree of freedom
in the fiducial metric is encoded only in $\hat{l}_{\mu\nu}$.

We can extend this kind of discussion to the higher order perturbations
and confirm the same result for them because the dynamics of the higher
order perturbations is also determined by the
functional form (structure) of the second order action, $S^{(2)}$. In
order to verify this statement, first, we consider the second order
perturbations,
\begin{eqnarray}
 \hat{h}_{\mu\nu} &=& \hat{h}_{\mu\nu}^{(1)} + \hat{h}_{\mu\nu}^{(2)},\label{h12}\\
 \hat{l}_{\mu\nu} &=& \hat{l}_{\mu\nu}^{(1)} + \hat{l}_{\mu\nu}^{(2)},\\
 \hat{\pi} &=& \hat{\pi}^{(1)} + \hat{\pi}^{(2)}.\label{pi12}
\end{eqnarray}
Here we regard only the second order perturbations as dynamical
variables and the linear perturbations are just (given) solutions of the
linear equations of motion. Since the full action can be written as
\begin{eqnarray}
 S &=& S^{(2)}[\hat{h},\hat{l},\hat{\pi}] + (\text{at least cubic order terms of }\hat{h},\hat{l},\hat{\pi} ),
\end{eqnarray}
the fourth order action can be schematically written as
\begin{eqnarray}
S^{(4)} &=& S^{(2)}[\hat{h}^{(2)},\hat{l}^{(2)},\hat{\pi}^{(2)}]
  + \int d^4 x \left(c_{nm}^{ijk} \Phi^{(2)}_i \partial^n \Phi^{(1)}_{j} \partial^m \Phi^{(1)}_{k}
  + (\text{terms without $\Phi^{(2)}$}) \right),
\end{eqnarray}
with some coefficients $c^{ijk}_{nm}$, where $\Phi^{(I)}_i $ represents
$I$-th order perturbations $\hat{h}^{(I)}_{\mu\nu} ,
\hat{l}^{(I)}_{\mu\nu},\hat{\pi}^{(I)}$.  The equations of motion of the
second order perturbations, which can be derived from fourth order
action, are given as
\begin{eqnarray}
 \frac{\delta S^{(2)}[\Phi^{(2)}]}{\delta \Phi^{(2)}_i } = - c^{ijk}_{nm} \partial^n \Phi^{(1)}_{j} \partial^m  \Phi^{(1)}_{k}.\label{eom2nd}
\end{eqnarray}
Since the left hand side is the same form as the equation of motion of
the linear perturbations, there is no higher time derivative term.
Since the dynamics of linear perturbation has already been determined by
the linear order equations of motion, the terms in right hand side of
Eq.~(\ref{eom2nd}) are just source terms. Then the higher time
derivative terms appearing in right hand side do not lead to the
Ostrogradsky instability. To be more concrete, such
higher time derivative terms can be reduced to lower derivatives ones by
use of the linear order equations of motion. The extension to arbitrary
higher order of perturbations is trivial. The equations of motion for
$N$-th order perturbations can be derived from $2N$-th order action, and
can be written as
\begin{eqnarray}
  \frac{\delta S^{(2)}[\Phi^{(N)}]}{\delta \Phi^{(N)}_i } = \text{source terms }.
\end{eqnarray}
The key observation is that there is no higher order derivative in the
left hand side while the right hand side includes higher order
derivatives but consists of up to the $(N-1)$-th perturbations, whose
dynamics has already been determined by the lower order equations of
motion. Thus, by the same discussion on the linear order perturbations,
the Ostrogradsky instability does not appear at any order of
perturbations in the decoupling limit. This is
none other than
the main conclusions of this note.

It is worth noticing that our discussion here is similar
to the one in the low-energy effective theory approach. In the
low-energy effective action, there appear higher derivatives of
low-energy degrees of freedom, e.g., as a consequence of integrating out
massive modes. However, those higher derivatives do not imply the
existence of ghosts, rather they provide the cutoff scale for the
derivative expansion. More practically, such higher derivatives are
eliminated by plugging the leading order equations of motion. See, e.g.,
Ref.~\cite{Weinberg:2008hq} for more details. The main difference in our
discussion is that we use the perturbative expansion based on the
smallness of the perturbations around the fixed metric, rather than the
derivative expansion. Just as the low-energy effective theory case,
higher order derivatives can be eliminated order by order by using the
lower order equations of motion, as long as the perturbations around the
background are small. Let us then close discussion by clarifying under
which conditions such a perturbative expansion is justified. In
$\Lambda_3$ decoupling limit, possible terms with $n\,(\geq 2)$-th order
in perturbations are as follows:
\begin{eqnarray}
\label{higher}
 \frac{1}{\Lambda_3^{3(n-2)}} \hat{h}(\bar{\nabla}^2 \hat{\pi})^{n-1},\ \frac{1}{\Lambda_3^{3(n-2)}}\bar{\nabla}^{2(n-1)-d}\hat{l}\bar{\nabla}^d\hat{\pi}^{n-1}
, \frac{1}{\Lambda_3^{3(n-2)}}\bar{\nabla}^{2(n-1)-d}\left(\frac{\bar{R}}{m^2}\right) \bar{\nabla}^d \hat{\pi}^n,
\end{eqnarray}
where $d$ is an integer which satisfies $d \leq 2(n-1)$. If we denote
the typical size of perturbations by $\epsilon$ and assume that
$\bar{R}\sim m^2$ for simplicity, the interaction terms~\eqref{higher}
do not dominate over the second order action (more precisely, the
$n\,(\geq 2)$-th order term dominates the $(n+1)$-th order term) as long as
\begin{align}
\bar{\nabla}^2\epsilon^2\gg \frac{1}{\Lambda_3^{3(n-2)}}\bar{\nabla}^{2(n-1)}\epsilon^n
\quad
\leftrightarrow
\quad
\Lambda_3^{3}\gg \bar{\nabla}^{2}\epsilon\,.
\label{eq:expand_cond}
\end{align}
This is the condition for the validity of our perturbative expansion and
we have shown that there are no BD ghosts in this regime. Notice that it
is naturally satisfied in the $\Lambda_3$ decoupling regime
because the scale of a derivative $\bar{\nabla}$,
denoted by $\Lambda$, can be at most of the order of $\Lambda_3$ in the
$\Lambda_3$ decoupling regime, which implies that $\Lambda_3^3 \ge
\Lambda^2\Lambda_3 \gg \Lambda^2 \epsilon$ for $\epsilon \ll \Lambda_3$.

\section{Conclusion}

We applied the St\"{u}ckelberg formalism of dRGT massive gravity with a
general fiducial metric established in Ref.~\cite{Gao:2014ula} to
bi-gravity. In the case of massive gravity, the decoupling limit of the
action includes the non-trivial coupling between the curvature of the
fiducial metric and the helicity-0 mode of St\"{u}ckelberg
fields, Eq.(\ref{L_unmix_4}). However, since the fiducial metric is
non-dynamical in massive gravity, such terms do not lead to the
dangerous BD ghost.

In the case of bi-gravity, where the fiducial metric is dynamical, one
may wonder if such terms would lead to the dangerous BD ghost. Then, we
have revisited this question.  First we derived the decoupling limit
action for the linear perturbation and confirmed that higher time
derivative terms do not appear, Eq.(\ref{S2}).  Next, we confirm that
the equations of motion of higher order perturbations are the same as
those of the linear order perturbations except for the source term
coming from the lower order perturbations, whose dynamics has already
been determined by the lower order equations of motion.  Then, by using
this result, we reconfirm that the Ostrogradsky instability (BD ghost)
does not appear for arbitrary order of the perturbations at the
decoupling limit in bi-gravity as long as perturbative expansion is
justified.

\section*{Acknowledgments}

This work was in part supported by a grant from Research Grants Council
of the Hong Kong Special Administrative Region [HKUST4/CRF/13G] (T.N.),
the JSPS Grant-in-Aid for Scientific Research Nos. 25287054 (M.Y.),
26610062 (M.Y.), the JSPS Grant-in-Aid for Scientific Research on
Innovative Areas No. 15H05888 (M.Y.), and the JSPS Research Fellowship
for Young Scientists, No. 26-11495 (D.Y.).

\section*{References}
\bibliographystyle{elsarticle-num}
 
\end{document}